\documentstyle[prb,twocolumn,epsf,floats,aps]{revtex}
\begin{document}
\draft

\twocolumn[\hsize\textwidth\columnwidth\hsize\csname @twocolumnfalse\endcsname

\title{Superconducting $d_{x^2-y^2} \pm id_{xy}$ phase glass}

\author{Patrik Henelius$^1$, A. V. Balatsky$^2$, J. R.
Schrieffer$^1$}

\address{$^1$ National High Magnetic Field Laboratory,
1800 East Paul
Dirac Dr., Tallahassee, Florida 32310\\
$^2$ Theoretical Division, Los Alamos National
Laboratory, Los Alamos, NM 87501}

\date{\today}

\maketitle

\begin{abstract}
We discuss the effects of magnetic impurities on d-wave
superconductors. We calculate the electron mediated RKKY interaction
between the impurity spins in a d-wave superconductor and find that it
decays as $r^{-3}$ at large distances. We argue that this interaction
leads to the formation of a spin glass at low temperature $T \ll
T_c$. It was previously shown that a local complex $\Delta^1 \sim
d_{xy}$ order parameter is induced around each impurity spin. We
consider the pair tunneling resulting in the Josephson interaction
between different patches of induced $d_{xy}$ order parameter.  Due to
the local coupling between impurity spins and the superconducting
order parameter the Josephson coupling favors a ferromagnetic phase at
low temperatures. The competition between the Josephson coupling and
the RKKY interaction gives rise to an interesting phase diagram. At
low impurity concentrations we find an unusual supercondcting phase
glass, where the impurity spins $S^z$ and $d_{xy}$ component are
disordered and yet the product of these two develops a true long range
order $\langle S^z \Delta^1\rangle$. This phase has no analog in
purely magnetic spin glasses and arises as a result of the direct
coupling of the impurity spin to the phase of $\Delta^1$.  At high
impurity concentrations it is possible that a ferromagnetic phase will
form.

\end{abstract}


\vskip2mm]

\section{INTRODUCTION}

It has been known for a long time that magnetic ions in a
superconductor can form a spin glass (SG) state.\cite{Dav1} There is
growing experimental evidence for the formation of a SG phase in the
high-$T_c$ materials once magnetic ions, such as Fe and Ni, are doped
into the system.\cite{SG1} Since magnetic ions interact with the
superconducting condensate it is natural to expect\cite{Kirk1} that
this interaction will cause frustration of the underlying condensate
and eventualy might lead to a superconducting glass or phase glass
(PG).

\begin{figure}[htb]
\centering
\epsfxsize=8cm
\leavevmode
\epsffile{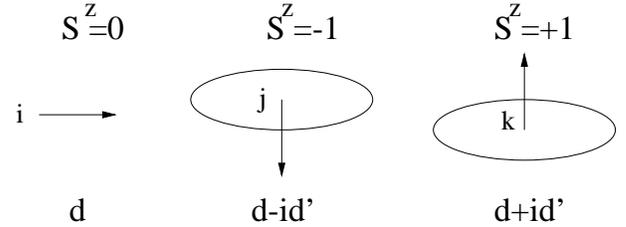}
\vskip0.5cm
\caption{Three impurity spins at sites $i$, $j$ and $k$ in a
two-dimensional $d_{x^2-y^2}$ superconductor. The patches around site
$j$ and $k$ indicate the induced $d'=d_{xy}$ component of the order
parameter. Note that there is no induced order parameter around spin
$i$, since the spin is pointing in the $xy$ plane.}
\label{patch}
\end{figure}

In all of the above discussions of the role of impurity spins it was
assumed that magnetic scattering frustrates and suppresses the
superconductivity.  There are additional physical effects that were
not addressed in previous work: namely how frustrated localized spins
in an unconventional superconductors can {\em distort} the condenstate
and produce {\em patches} of secondary components of the order
parameter near the localized impurity, see Fig.~\ref{patch}. It has
been shown that a magnetic impurity in a $d_{x^2-y^2}$ superconductor
with order parameter $\Delta^0$ induces a local complex $d_{xy}$
component of the order parameter which we designate
$\Delta^1$.\cite{B1} Locally near each impurity site, on the scale of
the coherence length $\xi$, there is a patch of $\Delta^0 + i\Delta^1$
order paramter for $S^z =+1$ and $\Delta^0 - i\Delta^1$ for $S^z =
-1$, depending on the sign of the impurity spin.  Hereafter we assume
classical spins with $S = 1$. This is a reasonable assumption taken
the fact that magnetic ions substituted into high-$T_c$
superconductors have a large spins, such as Ni (S=1).

Here we will explore the coupling between different patches due to the
Josephson coupling. Phase coherence between these patches of $\Delta^0
\pm i \Delta^1$ would lower the kinetic energy of the condensate and
would tend to align all the patches into {\em globally coherent}
$\Delta^0 +i\Delta^1$, or its conjugate. This would imply
ferromagnetic ordering of the impurity spins $S^z$. On the other hand,
the dipolar and RKKY spin-spin interaction terms, which we show are
mainly antiferromagnetic, would frustrate this ferromagnetic
order. In fact, we find  that the spin-spin interaction is
frustrating and produces a SG phase at $T<T_{SG}$ in the absence of a
coupling between the superconducting condenstate and spin degrees of
freedom.  When this coupling is considered the result depends on the
relative strength of the two interactions. We find that since the
spin-dependent part of the Josephson coupling is small the magnetic
susbsystem will drive the phase transition, at least at low and
intermediate impurity concentrations. Given that the impurity spins
form a SG phase we are led to the question how the phase of the
induced component $\pm i\Delta^1$ is affected by spin frustration.  We
find that the spin frustration in the SG phase will frustrate the
phase of the $d_{xy}$ component and a PG will form at low temperatures
$T< T_{PG} < T_{SG}$. More generally we will discuss the possible
phases of the coupled spin-phase model, and at higher impurity
concentrations we will consider the formation of a ferromagnetic
phase. When discussing glassy phases, we will consider states where
the spatial average $\langle S^z\rangle=0$.  However at any given
patch, $i$, the time averaged $\langle S_i^z\rangle_{\tau} \neq
0$. Similarly we will consider site and time averaged induced
component $\langle\Delta^1\rangle = 0$ and $\langle
\Delta^1_i\rangle_{\tau} \neq 0$ respectively.

The additional new order parameter we find relevant is the product
$\langle S^z\Delta^1\rangle$. The new order in this case arises from
the possibilty of having fully disordered phases
$\langle\Delta^1\rangle =\langle S^z\rangle = 0$ and still having the
{\em true long range order} in $\langle S^z\Delta^z\rangle\neq
0$. Physically it corresponds to phase locking of the patches to the
local value of impurity spin $S^z$. This order parameter has no analog
in purely magnetic spin glass systems and is a direct consequence of
the spin-phase order parameter coupling discussed above. In all of the
discussion hereafter we assume that the ``backbone'' order parameter
$\Delta^0$ has true long range order as it is robust at low
temperatures $T \ll T_c \sim 100 K$.

The plan of this paper is as follows: in the first section we explore
the effective spin-spin coupling and integrate out the superconducting
degrees of freedom by performing an RKYY calculation. The effective
spin model is found to give rise to a SG phase. In the second section
of the paper the Josephson coupling between different patches is
considered. It is found that the interaction favors a FM phase.  In
the last section we dicuss the PG phase, where the induced
superconducting order parameter is locked to the impurity
spin. Furthermore we will discuss the possibilities of para- and
ferromagnetic phases.

In related recent work\cite{var} Simon and Varma treat the magnetic
impurity problem by a variational approach. They concentrate on the
single impurity, but conclude by arguing that the formation of a
ferromagnetic phase is unlikely and that a spin glass state is much
more likely.

\section{EFFECTIVE SPIN MODEL}
In this section we will consider the effective spin model that
describes the interaction between impurity spins surrounded by patches
of induced complex $d_{xy}$ order parameter, as shown in
Fig.~\ref{patch}. The most relevant sources of interaction between the
impurity spins are the electron-mediated RKKY-interaction and the
direct dipolar magnetic interaction.  The standard RKKY
interaction\cite{rkky} is mediated through an interaction of the form
$JS\cdot I$, where $S$ referes to the spin of the conduction electron,
and $I$ denotes the spin of the impurity. Of particular interest to
this work, however, is another term that is directly related to the
induced order parameter. The $L\cdot I$ interaction, where $L$ refers
to the angular momentum of a conduction electron, scatters an electron
into the complex $d_{xy}$ phase. A second order RKKY calculation where
the interaction potential is taken to be of the from $L\cdot I$ will
therefore be relevant in this context. The effective spin Hamiltonian
is thus of the form
\begin{equation}
H= H^M + H^{S\cdot I} + H^{L\cdot I},
\end{equation}
where the dipolar term $H^M$ and the electron mediated interactions
$H^{S\cdot I}$ and $H^{L\cdot I}$ are given by
\begin{eqnarray}
H^M&=&\sum_{ij}\frac{1}{r_{ij}^3}[S_i\cdot S_j-3(S_i\cdot \hat r_{ij})
(S_j\cdot \hat r_{ij})]\\ 
H^{S\cdot I}&=&\sum_{ij} J_{ij}^{S\cdot I}S_i\cdot S_j\\ 
H^{L\cdot I} &=&\sum_{ij} J_{ij}^{L\cdot I}S_i^z S_j^z
\end{eqnarray}

The size of the dipolar term for two spins separated by $1$ \AA can be
estimated to be about $10^{-4}$ eV, while the RKKY interaction depends
strongly on the coupling between the conduction electrons and the
impurity spin. For Mn ions in alloys is has been estimated to be on
the order of $10^{-1}$ eV.\cite{smith} We are not aware of direct
estimates of the RKKY interaction in high-$T_c$ materials, but it is
likely to be greater than the dipolar interaction. The spatial decay
of the last two terms are explicitly calculated in the appendix and
are of the form
\begin{eqnarray}
H^{L\cdot I}&\sim & \frac{2J^2I_1^zI_2^zk_F^2}{ vr^3}[0.37+0.33\sin(2Kr)]\\
H^{S\cdot I}&\sim & \frac{J^2I_1\cdot I_2}{2\pi vr^3}[0.37+0.33\sin(2Kr)]
\end{eqnarray}
where $v$ is the gap velocity and K is a momentum cut-off.  The RKKY
terms thus share a cubic decay with the dipolar interaction, but are
found to be completely antiferromagnetic.  However, independently of
the specific form of the terms, general consideration tells us that as
long as there are sufficient amounts of disorder and frustration
present the SG phase should be realized.\cite{bind} The impurity spins
we are considering are randomly distributed, and antiferromagnetic
interactions will therefore lead to frustration. So as long as the
effective interactions are not primarily ferromagnetic the model
should have a SG groundstate. The dipolar interaction is
antiferromagnetic in the $z$-component of the spins in an $xy$ plane,
and ferromagnetic in the plane. As mentioned above the RKKY terms are
completely antiferromagnetic. We therefore conclude that the effective
spin model, consisting of the combined dipolar and RKKY terms,
exhibits a spin glass phase at low temperatures. For large enough
quantum fluctuations it would also be possible to realize a quantum
paramagnet phase, where any freezing is destroyed by quantum
fluctuations. We focus here on large classical spins and assume that
quantum fluctuations are negligible. In principle similar
considerations can be given for $S=1/2$ impurity spins, in which case
the paramagnetic phase could be realized.

Note in particular that the impurity spins would form a spin glass
even without the $H^{L\cdot I}$ term, and that this term is second
order in $\Delta^1 \ll \Delta^0$ . This is the motivation why we have
first considered the effective spin model and determined the spin
configuration.  So although the spin and superconducting degrees of
freedom are connected, as we will see in the next section, we will
there take the spin configuration as given, and determine the phase of
the superconducting order parameter based on this.  Note also that the
importance of the dipolar term may increase as the $\Delta_1$
component is induced, since this will open up a fully gapped state
which will tend to exponentially decrease the long-range RKKY
interaction. The short-range RKKY interaction should not be
significantly affected by the opening of the superconducting gap.

\section{JOSEPHSON COUPLING}

In this section we will consider the Josephson coupling between the
different patches of induced order parameter. Let us first consider
the order parameter around an impurity spin
\begin{equation}
\Psi_i=(\Delta_i^0+e^{i\frac{\pi}{2}S_i^z}\Delta_i^1)e^{i\theta_i},
\label{loc}
\end{equation}
where the real parameter $\Delta^0$ denotes the $d_{x^2-y^2}$ order
parameter, and the likewise real quantity $\Delta^1$ denotes the
induced $d_{xy}$ component. Hereafter we assume that $\Delta^0$ is a
robust order parameter that develops true long-range order at $T<T_c
\simeq 90$ K and remains ordered in all of the phases we discuss.  The
impurity spin $S_i^z$ uniquely determines the relative phase of the
induced order parameter to be $+\pi/2$ for $S^z=+1$ and $-\pi/2$ for
$S^z=-1$. The main result of the previous section was that the
impurity spins, considered independently of the coupling to the
induced order parameter, will exhibit a spin glass phase at low
temperatures. Let us next consider what effects the inter-patch
Josephson interaction may have.

In order to address this question we examine the Josephson coupling
between different patches given by
\begin{equation}
H^J=-\sum_{ij} |I_{ij}|(\Psi_i^*\Psi_j + \Psi_i\Psi_j^*),
\end{equation}
Using the local order parameter Eq.~(\ref{loc}) this leads to an
interaction of the form
\begin{equation}
H^J=-2\sum_{ij} |I_{ij}|[(\Delta^0)^2+(\Delta^1)^2S_i^zS_j^z]
\cos(\theta_i-\theta_j)
\end{equation}

We have here assumed that $\Delta_i^0=\Delta_j^0=\Delta^0$, and
likewise for $\Delta_1$. The first term, which is zeroth order in
$\Delta_1$, wants to align the phase of $\Delta^0_i$ at different
patches by setting $\theta_i=\theta_j$. The second term wants to align
the impurity spin in an ferromagnetic phase. The Josephson coupling
thus favors a ferromagnetic spin configuration, while the effective
RKKY and dipolar spin model favors a spin glass. In the next section
we will consider some possible outcomes of this competition.

We would, however, like to point out that the RKKY and Josephson
effects are not as independent as they may at first seem. Both are
mediated by electrons and while the dominant part of the RKKY
interaction in the superconducting state is zeroth order in $\Delta^1$
there is a part that is second order in $\Delta^1$, corresponding to
the Josephson coupling.  The Josephson interaction physically
expresses an effect arising from electron pair tunneling, while the GG
part of the RKKY interaction expresses an electron-hole channel. Here
G and F are the normal and anomalous propagators in superconducting
state.  The FF part of the RKKY interaction is closer to the Josephson
interaction since it expresses an electron-electron process.
Furthermore, there is a second order contribution to the RKKY
interaction which we have not evaluated in this work. This contains
two explicit $d_{xy}$ propagators, with two exchange interactions at
the impurity spins. This part of the interaction should be similar to
the Josephson interaction since explicit information about the
patches, such as spatial decay, should be contained in the
propagators. The physical effects of this term should, however, be
included in the Josephson coupling considered above.

\section{POSSIBLE PHASES}

In this section we will consider the different phases that could occur
as a result of the competition between the RKKY and the Josephson
terms. First we will enumerate the different phases and thereafter we
will discuss them in some more detail.

The simplest phase is a ferromagnetic (FM) phase, favored by the
Josephson coupling of different patches. The FM phase is characterized
by a finite spatial average of the magnetization and the induced
phase; $\langle S^z\rangle\neq 0$ and $\langle \Delta_1\rangle\neq 0$.

The phase favored by the effective spin Hamiltonian, on the other
hand, is a SG phase, where the spatial average of the magnetization
vanishes
\begin{equation}
\langle S^z \rangle =0,
\end{equation}
but the time averaged local magnetization remains finite;
\begin{equation}
\langle S^z_i \rangle_\tau \neq 0.
\end{equation}
Assuming that the phase of $\Delta^1$ is determined by the impurity
spin this gives rise to similar ordering for the induced phase;
$\langle\Delta^1\rangle =0$ and $\langle \Delta^1 \rangle_\tau \neq
0$.  In this case the combination $S^z\Delta^1$ will also have a
non-vanishing spatial average
\begin{equation}
\langle S^z\Delta^1\rangle \neq 0
\end{equation}
This particular order parameter is unique for the coupling of SG and
SC degrees of freedom and is not present in a purely magnetic
system. It describes the phase-locking of the superconducting phase
and the impurity spin.\cite{abrahams} There also exists the
possibility that the induced order parameter $\Delta^1$ does not
phase-lock with the impurity spin, but prefers to vary over length
scales greater than the inter-patch distance.\cite{joyn} In that case
the spatial average $\langle S^z\Delta^1\rangle$ would vanish, but the
local time average $\langle S^z_i\Delta^1_i\rangle_{\tau} $ would
still remain finite.

If quantum fluctuations are strong enough, then it would also be
possible to realize a paramagnetic spin phase. In this case also the
time averages $\langle S^z\rangle_{\tau}$ and $\langle
\Delta^1\rangle_{\tau}$ would vanish, and only local $\langle
S^z\Delta^1\rangle $ would remain finite, for as long as the
phase-locking is maintained. In this work we will, however, neglect
this possibility since we focus on large classical spins.

After having considered the different options let us consider the
interplay between the different terms. If there is a low concentration
of impurity spins (the patches do not overlap), then the ferromagnetic
Josephson term is bound to be exponentially small, and the ground
state should be a spin glass, characterized by a non-vanishing spatial
average $\langle S^z\Delta^1\rangle $. If the impurity concentration
becomes larger, but not large enough to kill the $\Delta^0$
superconductivity, then there may be a region where the ferromagnetic
Josephson terms are predominant and the FM phase is formed. The FM
phase was considered in detail in a previous Ginzburg-Landau
description \cite{B1}.

\begin{figure}[htb]
\centering
\epsfxsize=6cm
\leavevmode
\epsffile{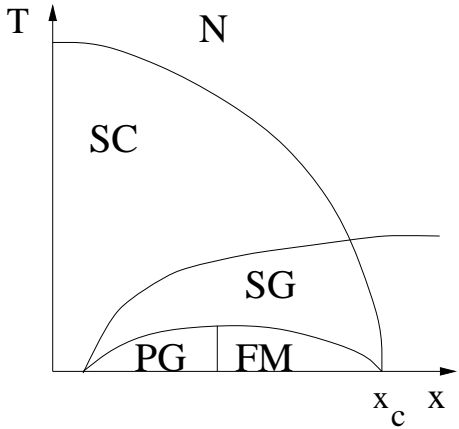}
\vskip0.5cm
\caption{Phase diagram for the impurity doped d-wave superconductor as
function of impurity doping (x) and temperature (T).  The labeled
diagrams show the normal (N) phase, the $d_{x^2-y^2}$ superconducting
(SC) phase, the spin-glass (SG) phase, the superconducting
$d_{x^2-y^2}+id_{xy}$ phase-glass (PG) phase and the ferromagnetic
(FM) phase. The SC phase is suppressed by disorder due to
pair-breaking, while the SG phase is independent of the SC order
parameter and will persist in the N phase. The PG phase is induced by
the SC and SG phases and exists only within these phases. The possible
FM phase is a result of strong Josephson coupling when the impurity
concentration is large. It is strongly suppressed close to $x_c$ due
to the suppression of the superconducting order.}
\label{phase}
\end{figure}

In Fig.~\ref{phase} we present a phase diagram as a function of
impurity concentration and temperature. Note that this PG phase is
different from previously proposed superconducting phase glasses in
that the glassy behavior is only displayed in the induced component of
the order parameter, and not in the robust $d_{x^2-y^2}$
part. Considering the phase diagram we note that the disorder
suppresses the critical temperature for the superconductor-metal
insulator, as is well known. The impurity spins form a spin-glass
phase at low temperatures, and this phase is independent of the
electronic order parameter and persists also in the metallic
phase. The PG phase is induced by the SG and SC phases and hence it
must only exist within the boundaries of these two phases.  The
possible FM phase is induced by a large impurity concentration.  As
the $d_{x^2-y^2}$ superconducting order parameter gets suppressed by
disorder the induced component will also vanish, and hence the FM
phase gets strongly suppressed as we approach $x_c$.

Experimentally the proposed phase locked state can be observed in
scanning tunneling microscope measurements, where the particle spectrum
would develop a full gap near impurity site even though the phase of
$\Delta^1$ remains uncertain\cite{B1}. The ac magnetic susceptibility in the
superconducting state also should show features upon crossing the SG
and PG lines. Another experiment, that would be sensitive to the
appearance of the $\langle S^z \Delta_1 \rangle$, would be the
penetration depth that would become exponential $\delta 1/\lambda^2
\propto \exp[-|\Delta^1|/T]$ in the PG and FM phases.

\section{CONCLUSION} 

We have examined the role of magnetic impurities in a d-wave
superconductors. In particular we have studied the effective impurity
spin model arising from electron mediated RKKY and magnetic dipolar
terms and argued that these terms lead to a SG phase at low
temperatures.. Furthermore we have analyzed the coupling between the
spin- and superconductor order parameter that arises from the
Josephson interaction of patches of induced order parameter around the
impurity spins. The Josephson interaction favors a FM phase. At high
impurity concentrations the FM phase may be realized, while at low
concentrations the SG phase would be prefered. Due to the coupling
between the spin and superconducting order parameters the SG phase
induces a superconducting PG at low temperatures. The glassy behavior
is a property of the induced $d_{xy}$ component of the order
parameter, while the primary component $d_{x^2-y^2}$ is assumed
robust. The superconducting PG phase is characterized by an order
parameter of the form $\langle S^z\Delta^1\rangle$, which describes
the phase-locking of the induced order parameter and the local
impurity spin.  In addition to the PG and FM phases we have also
discussed a possible paramagnetic phases.

We are grateful to D. Agterberg, N. Bonesteel, M. Graf and I. Martin,
for useful discussions.

This work was supported by US DOE and NSF Grant No. DMR-9629987.

\begin{appendix}

\section{The RKKY interaction}
The RKKY interaction describes a second-order process, where an
electron of momentum k interacts with an impurity spin at $r=R_1$, is
scattered to a state with momentum q and interacts with a another
impurity spin at $r=R_2$, where it is scattered back into the original
state. This process is described by the following diagram:

\begin{figure}[htb]
\centering
\epsfxsize=6cm
\leavevmode
\epsffile{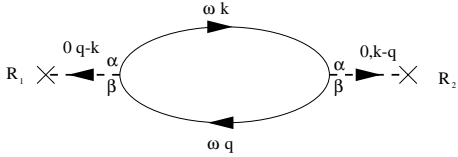}
\vskip0.5cm
\caption{Second-order RKKY interaction.}
\label{fig:dia2}
\end{figure}

Using Feynman rules we get the following expression for the effective
Hamiltonian
\begin{eqnarray}
-iH&=&(-1)\int {d^dk\over (2\pi)^d}\int {d^dq\over (2\pi)^d}\int {d\omega
       \over (2\pi)}\times\nonumber\\
      & & iG^0(\omega,k)(-iV^{R^1}_{k\alpha,q\beta})
       iG^0(\omega,q)(-iV^{R^2}_{q\beta,k\alpha})\\
iH&=&\int {d^dk\over (2\pi)^d}\int {d^dq\over (2\pi)^d}\int {d\omega
       \over (2\pi)}\times\nonumber\\
    & & G^0(\omega,k)V^{R^1}_{k\alpha,q\beta}
       G^0(\omega,q)V^{R^2}_{q\beta,k\alpha}
	\label{effi}
\end{eqnarray}

Using the Nambu formalism for the superconductor this expression
transforms to
\begin{eqnarray}
H&=&\int {d^2k\over (2\pi)^2}\int {d^2q\over (2\pi)^2} \int 
{d\omega \over (2\pi i)}\times \\
& &\mbox{Tr}\left\{[G(\omega,k)][V^{R^1}_{k,q}\tau_0]
[G(\omega,q)][V^{R^2}_{q,k}\tau_0]\right\},\nonumber
\end{eqnarray}
where 
\begin{eqnarray}
G(\omega,k)&=&\left[\begin{array}{cc}G_{11}(\omega,k) & F(\omega,k) \\
	     F(\omega,k) & G_{22}(\omega,k)\end{array}\right]\\
\tau_0&=& \left[\begin{array}{cc} 1&0\\ 0&1 \end{array}\right]\\
F(\omega,k)&=&\frac{\Delta_k}{\omega^2-E_k^2+i\delta}\\
G_{11}(\omega,k)&=&\frac{\omega+\epsilon_k}{\omega^2-E_k^2+i\delta}\\
G_{22}(\omega,k)&=&\frac{\omega-\epsilon_k}{\omega^2-E_k^2+i\delta},
\end{eqnarray}
where $E_k=\sqrt{\epsilon_k^2+\Delta_k^2}$. For a tight
binding model $\epsilon_k=-t\left[\cos(k_xa)+\cos(k_ya)\right]$,
and the $d_{x^2-y^2}$ gap is given by $\Delta_k=\Delta_0
\left[\cos(k_xa)-\cos(k_ya)\right]$.

Next we will consider a few specific forms of the interaction
$V^{R}_{k\alpha,q\beta}$, which is the Fourier transform of the
electron-impurity spin interaction. For free electrons
$|k\rangle=e^{ikx}$ and we get
\begin{equation}
V^R_{k\alpha,q\beta}=\langle k\alpha|V^R|q\beta\rangle =\int d^dx
                       e^{-i(k-q)x}\langle\alpha|V^R|\beta\rangle
\end{equation}
Let us first consider a contact potential of the form
$V^R=J\delta(x-R) S\cdot I$, where $S$ denotes the electron spin and
$I$ the impurity spin. This results in
\begin{equation}
V^R_{k\alpha,q\beta}=Je^{-i(k-q)R}\langle\alpha|S\cdot I|\beta\rangle
\end{equation}

Next we will consider a $L\cdot I$ interaction, where $L$ is the
angular momentum of the electron, and $I$ is the spin of the impurity
at location $R$. The electron moves relative to the impurity spin,
which sees a magnetic field of the form $B \propto {L\over |x-R|^3}$,
where $x$ is the location of the electron. We will consider two
spatial dimensions, and since the angular momentum of the electron
will have only a $z$-component we consider an interaction of the form
\begin{equation}
V^R=J{L\cdot I\over |x-R|^3}=J{L^zI^z\over|x-R|^3},
\end{equation}
where $L=(x-R)\times p =-i(x-R)\times\nabla_x$. We get the following
expression for the matrix element:
\begin{equation}
V^R_{k\alpha,q\beta}=J \int d^2x e^{-ikx}{[-i(x-R)\times\nabla_x]^z
\over|x-R|^3}I^ze^{iqx}\langle\alpha|\beta\rangle
\end{equation}
After performing the integrals we arrive at
\begin{equation}
V^R_{k\alpha,q\beta}=-2\pi iJI^ze^{-i(k-q)R}
\frac{(k\times q)^z}{|k-q|}\delta_{\alpha,\beta}
\label{li}
\end{equation}

We are now in a position to evaluate the effective interaction
\begin{eqnarray}
H&=&-i\int {d^2k\over (2\pi)^2}\int {d^2q\over (2\pi)^2} \int {d\omega
\over (2\pi)} V^{R^1}_{k,q} V^{R^2}_{q,k}\times\nonumber\\
&&[G_{11}(\omega,k)G_{11}(\omega,q)+
G_{22}(\omega,k)G_{22}(\omega,q)\nonumber\\ & & +
2F(\omega,k)F(\omega,q)].
\end{eqnarray}
We will begin with the frequency integral for the FF
contribution
\begin{equation}
I_F=\int {d\omega \over (2\pi)} 2F(\omega,k)F(\omega,q).
\end{equation}
There are poles at $\omega=\pm \sqrt{E_k^2-i\delta}=\mp E_k\pm
i\delta$.  Closing the integral in the upper half plane leads to
contributing poles at $\omega=- E_k+i\delta$ and $\omega=-
E_q+i\delta$. It follows that
\begin{equation}
I_F=i\frac{\Delta_k\Delta_q}{E_kE_q(E_k+E_q)}
\end{equation}
Next we will consider the GG contribution
\begin{equation}
I_G=\int {d\omega \over (2\pi)} G_{11}(\omega,k)G_{11}(\omega,q)+
G_{22}(\omega,k)G_{22}(\omega,q)
\end{equation}
Proceeding as above it follows that
\begin{equation}
I_G=i\frac{(\epsilon_k\epsilon_q-E_kE_q)}{E_kE_q(E_k+E_q)}.
\end{equation}
We will start by considering an interaction of the $L\cot I$ 
kind. We then have
\begin{equation}
V^{R^1}_{k,q} V^{R^2}_{q,k}=
4\pi^2J^2I_1^zI_2^ze^{-i(k-q)r}\left|\frac{k\times q}{k-q}\right|^2,
\end{equation}
where $r=R_1-R_2$. For the effective interaction we then get
\begin{eqnarray}
H&=&\frac{J^2I_1^zI_2^z}{4\pi^2}\int d^2k\int d^2q 
e^{-i(k-q)r}\times\nonumber\\
& &\left|\frac{k\times q}{k-q}\right|^2
\frac{(\Delta_k\Delta_q+\epsilon_k\epsilon_q-E_kE_q)}{E_kE_q(E_k+E_q)}
\end{eqnarray}

\begin{figure}[htb]
\centering
\epsfxsize=5cm
\leavevmode
\epsffile{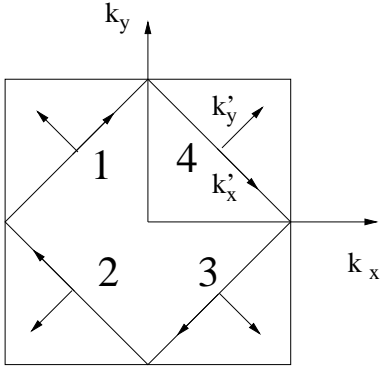}
\vskip0.5cm
\caption{Node notation.}
\label{fignode}
\end{figure}

In order to solve the above integral we introduce the nodal point
approximation.\cite{lee} The dominant contribution to the integral should come
from each nodal point, and we perform a rotation and translation to
transform the origin to the nodal points, with the $x$-axis along the
tight binding Fermi surface, see Fig.~\ref{fignode}.

The transformation is given by
\begin{equation}
\left(\begin{array}{c}k_x' \\k_y'\end{array}\right)
=\left(\begin{array}{cc} \cos\theta & \sin\theta\\-\sin\theta & \cos\theta
\end{array}\right) \left(\begin{array}{c}k_x \\k_y\end{array}\right)
-\left(\begin{array}{c}0 \\k_F\end{array}\right),
\end{equation}
where $k_F=\pi/\sqrt{2}a$ and $\theta =\{\pi/4,3\pi/4,5\pi/4,7\pi/4\}$
for the four nodes.

Next we need to apply these transformations to all quantities in the
effective interaction. The energy of a tight-binding model,
$\epsilon_k=-t\left[\cos(k_xa)+\cos(k_ya)\right]$, will transform
according to $\epsilon_k=v_Fk_y'$ for all nodes, and the gap function,
$\Delta_k=\Delta_0\left[\cos(k_xa)-\cos(k_ya)\right]$ will transform
according to $\Delta_k=\pm v_{\Delta}k'_x$, with a positive sign for
nodes 1 and 3, and a negative sign for nodes 2 and 4. Assuming an
isotropic superconductor $v=v_F=v_{\Delta}$ this leads to $E_k=vk'$.

The rotation and translation cannot affect $|k-q|$, and
therefore $|k-q|=|k'-q'|$. Furthermore 
\begin{eqnarray}
(k\times q)^z&=&k_xq_y-k_yq_x \cr
	&=& k_x'q_y'-k_y'q_x'+k_F(k_x'-q_x')
\end{eqnarray}
for all the four nodes. The second term is linear in the momentum and
will be retained. The final term to be transformed, $(k-q)r$, is
considered next. This term depends on the direction of $r$, and
different nodes will give different contributions. Assume that $r$ is
fixed in a direction $\theta_r$, and the contribution to the effective
interaction from node 1 has been calculated. Since all other terms are
identical, the contribution from the other nodes must be given by
substituting $\{ \theta_r+\pi/2, \theta_r+2\pi/2, \theta_r+3\pi/2\}$
for $\theta_r$ in the expression obtained for node 1. Therefore we
will look at how the transformation is done for node 1, and the
results for the other nodes will follow. For node 1 we get
\begin{eqnarray}
(k-q)r&=& r_x(k_x-q_x)+r_y(k_y-q_y)\cr
&=& \frac{r|k'-q'|}{\sqrt{2}}[\cos\theta_{k'-q'}(\cos\theta_r+\sin\theta_r)\cr
& &+\sin\theta_{k'-q'}(-\cos\theta_r+\sin\theta_r)]
\end{eqnarray}
As a summary we have thus arrived at the following transformations
\begin{eqnarray}
\epsilon_k&=&vk_y'\cr
\Delta_k&=&vk_x'\cr
E_k&=&vk'\cr
|k-q|&=&|k'-q'|\cr
(k\times q)^z&=&k_F(k'-q')^x\cr
(k-q)r&=&\frac{r|k'-q'|}{\sqrt{2}}[\cos\theta_{k'-q'}(\cos\theta_r+\sin\theta_r)
\cr
& &+\sin\theta_{k'-q'}(-\cos\theta_r+\sin\theta_r)]\nonumber
\end{eqnarray}
Dropping the primes, and using these results we can thus linearize the
effective interaction around the nodes:
\begin{eqnarray}
H&=&\frac{J^2I_1^zI_2^z}{4\pi^2}\int d^2k\int d^2q 
e^{-i(k-q)r}\times\cr
& &\frac{k_F^2|k-q|^2\cos^2\theta_{k-q}}{|k-q|^2}
\frac{vk_xvq_x+vk_yvq_y-vkvq}{vkvq(vk+vq)}
\end{eqnarray}
We start by integrating over the angles. First we fix the relative
angle $\theta$ between $k$ and $q$ and integrate over $\theta_k$,
thereafter we integrate over $\theta$.  Integrating over $\theta_k$ we
get
\begin{eqnarray}
H&=&\frac{J^2I_1^zI_2^zk_F^2}{4\pi^2v}\int dk\int dq \frac{kq}{k+q}
\int d\theta (\cos\theta -1)\times\cr
& &\pi\big[J_0(r|k-q|)-\sin(2\theta_r)J_2(r|k-q|)\big]
\end{eqnarray}
The angular dependence will, however, vanish, since summing up the
contributions from the four nodes gives us
\begin{eqnarray}
& &\sin(2\theta_r)+\sin(2(\theta_r+\pi/2))\cr
& &+\sin(2(\theta_r+\pi)) +\sin(2(\theta_r+3\pi/2))]=0,
\end{eqnarray}
and this tells us that the effective interaction is isotropic, even
though the gap is anisotropic.  Integrating out the relative angle we
find
\begin{eqnarray}
H&=& \frac{2J^2I_1^zI_2^zk_F^2}{v}\int dk\int dq\quad\times\cr
& &\frac{kq}{k+q}
\left[-J_0(kr)J_0(qr)+J_1(kr)J_1(qr)\right]
\end{eqnarray}
This integral is oscillatory, and we introduce a momentum cut-off $K$
and make the integration variables dimensionless by letting
$k\rightarrow rk$ and $q\rightarrow rq$:
\begin{eqnarray}
H&=&\frac{2J^2I_1^zI_2^zk_F^2}{vr^3}
\int_0^{Kr} dk\int_0^{Kr} dq\\
& &\frac{kq}{k+q}\left[-J_0(k)J_0(q)
+J_1(k)J_1(q)\right],
\end{eqnarray}
This integral can be solved numerically, and the behavior for large
$r$ is given by
\begin{equation}
H\sim \frac{2J^2I_1^zI_2^zk_F^2}{vr^3}[0.37+0.33\sin(2Kr)].
\end{equation}
This results represents an anti-ferromagnetic $r^{-3}$ part that is
independent of the cut-off, and an oscillating $\sin(2Kr)r^{-3}$ part.
Due to the relative sizes of the two terms the interaction is always
positive and hence completely anti-ferromagnetic. The $S\cdot I$
interaction will lead to a very similar result, differing only in the
prefactor.  This can be seen, since the result, after performing the
first angular integral will be given by
\begin{eqnarray}
H&=&\frac{J^2I_1\cdot I_2}{4\pi^3v}\int dk\int dq \frac{kq}{k+q}
\times \cr
& &\int d\theta (\cos\theta -1) \pi J_0(r|k-q|)
\end{eqnarray}
and the only difference compared to the $L\cdot I$ interaction is the
prefactor. The final result for the $S\cdot I$ coupling will be
\begin{equation}
H^{S\cdot I}\sim  \frac{J^2I_1\cdot I_2}{2\pi vr^3}[0.37+0.33\sin(2Kr)]
\end{equation}
So the two electron-spin couplings lead to the same functional form of
the RKKY interaction. We have therefore showed that both the $S\cdot
I$ and $L\cdot I$ interactions give rise to an effective
anti-ferromagnetic model. We have used the nodal approximation and
assumed that the superconductor is isotropic $v_F=v_{\Delta}$, and
these approximations may change the final result somewhat, but it
appears unlikely that they would make the effective spin-spin
interactions predominantly ferromagnetic, and therefore the randomly
distributed spins will form a spin-glass phase, as discussed in the
main part of the paper.  \vfill \break
\end{appendix}

\end{document}